\documentclass[11pt,twoside]{article}

\usepackage{asp2004}
\usepackage{epsf}
\usepackage{psfig}
\usepackage{lscape}
\usepackage{graphicx}

\begin{document}
\title{Detecting Clusters of Galaxies at High Redshift with the {\em Spitzer} Space
Telescope}   %%% Fill in title
\author{Gillian Wilson, Adam Muzzin, Mark Lacy  \& the FLS Team}   %%% Fill in author names
\affil{Spitzer Science Center, University of Toronto, Spitzer Science Center}    %%% Fill in author affiliations

\begin{abstract} %%% Abstract to run on from here.

We present an infrared adaptation of the Cluster Red-Sequence method.
We modify the two filter technique of Gladders \& Yee (2000)
to identify clusters based on their $R-[3.6]$ color. We apply
the technique to the 4 degree$^2$ {\em Spitzer} First Look Survey and 
detect 123 clusters spanning the redshift range $0.09<z<1.4$.
Our results demonstrate that the  {\em Spitzer} Space Telescope will
play an important role in the discovery of large samples of high redshift
galaxy clusters.

\end{abstract}

%%% MAIN BODY OF TEXT GOES HERE. CONSULT "INSTRUCTIONS FOR AUTHORS USING
%%% LATEX2E MARKUP", SECTIONS 2.3-2.6 FOR HELP WITH EQUATIONS, FIGURES,
%%% AND TABLES.

\section{Introduction}   %%% Top level section head (remove "%" symbol)

\begin{figure}[!ht]
\begin{center}
\scalebox{0.7}{
 \includegraphics[24,19][510,396]{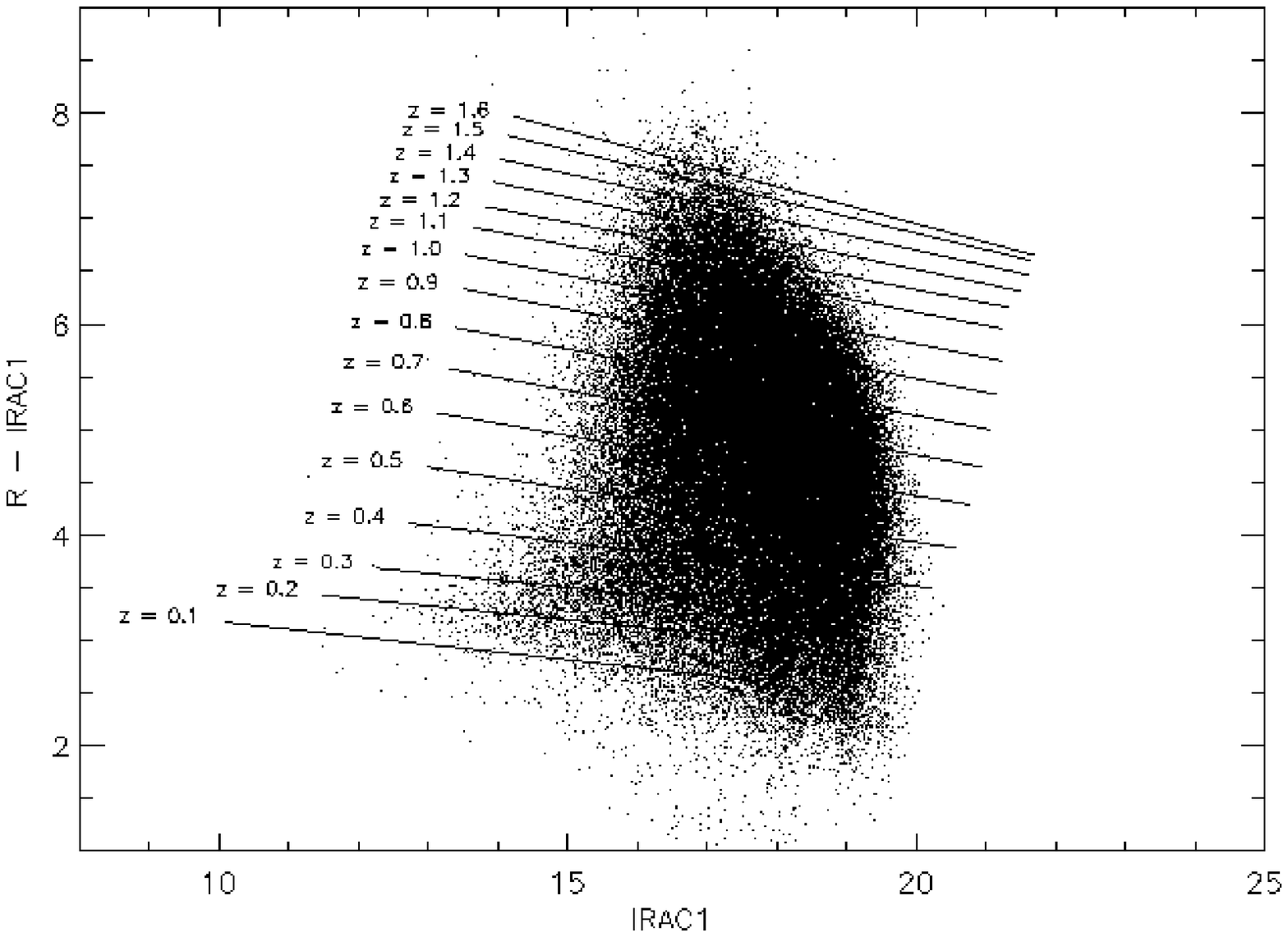}		
}
\caption{Color-magnitude diagram for all galaxies in the First Look Survey. 
The predicted $R-[3.6]$ color of an early type
galaxy at each redshift is shown  by the solid lines.
}
 \label{fig:CMR}
\end{center}

\begin{center}
\scalebox{0.7}{
 \includegraphics[80,345][565,700]{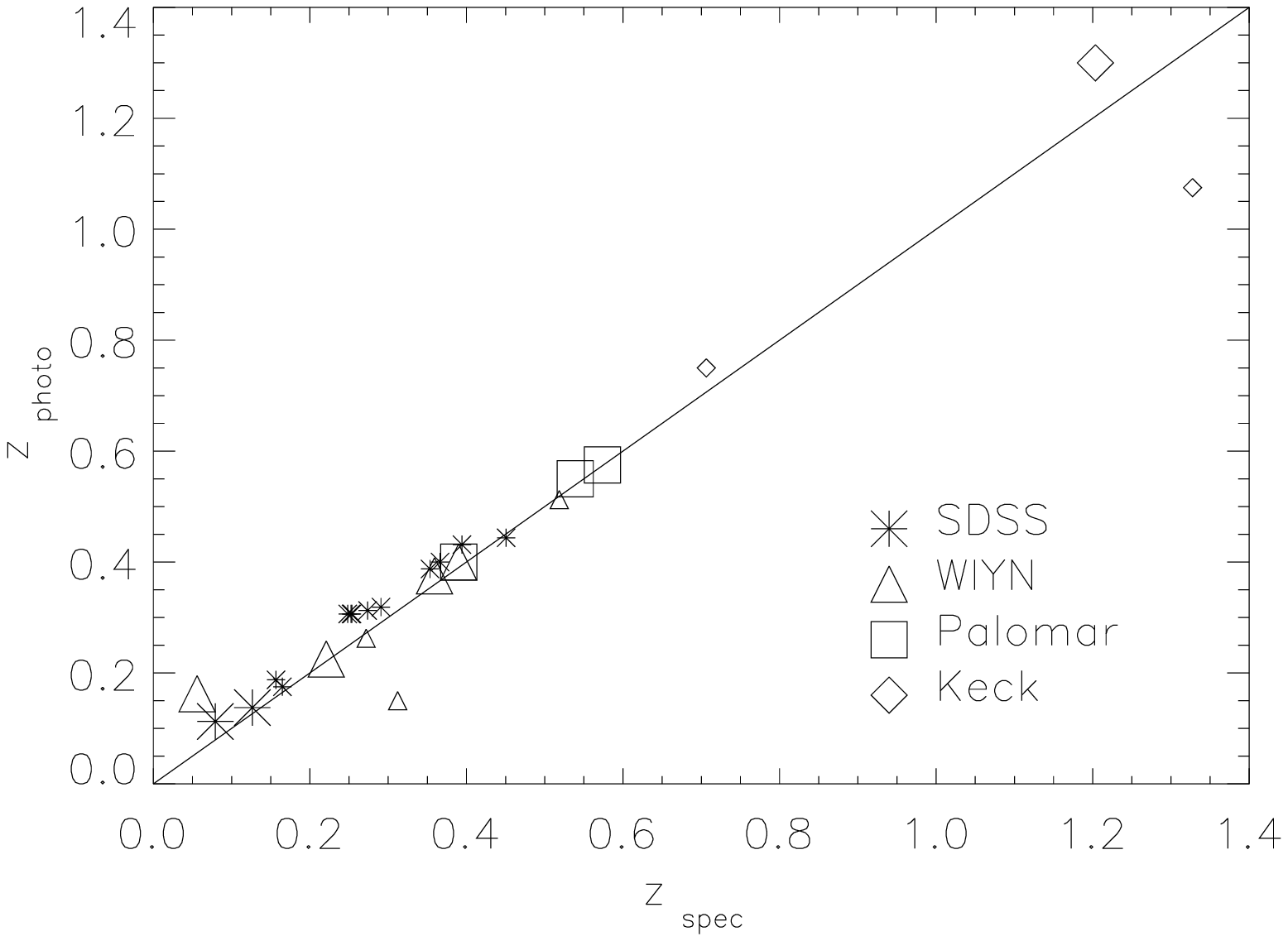}	
}
\caption{Plot of photometric versus spectroscopic redshifts. 
The large symbols indicate
clusters for which more than one spectroscopic redshift was measured.
The one-color $R-[3.6]$ photometric redshifts are in excellent agreement
with the spectroscopic redshifts with an r.m.s. dispersion of $\delta z = 0.07$.
}
 \label{fig:photoz}
\end{center}
\end{figure}

Early type (elliptical) galaxies in any given cluster have a 
similar color almost independent of their magnitude. If a color-magnitude
diagram is constructed using two filters that span the rest-frame $4000\AA$
 break, cluster early-types form a distinctive red sequence comprised of the brightest, reddest galaxies 
at a given redshift. Clusters can therefore be detected as overdensities
in a simultaneous projection of galaxy angular position, color and magnitude.
Furthermore, the color of the red sequence itself provides a precise estimate
of the redshift of the detected cluster (Gladders \& Yee 2000). The
Cluster Red-Sequence (CRS) technique is extremely insensitive to
projection effects, because random projections do not exhibit the
necessary red sequence signature in the color-magnitude plane. Cluster
surveys employing the CRS algorithm have traditionally been carried out using
an $R-z^{\prime}$ combination (Gladders \& Yee 2005).

Applying the CRS technique to higher redshift is an obvious next step. 
At $z =1.2$, the z$^\prime$ filter is no longer redward of the rest-frame $4000\AA$ break.  Therefore, to detect early-types at higher redshift, a redder filter must be employed. Covering large areas of sky in the NIR from the ground is a 
daunting proposition because of the bright NIR sky background and current dearth of
large format NIR cameras. 
The {\em Spitzer} Space Telescope offers a promising alternative.
Here we present results from the First Look Survey (FLS) utilizing an $R-[3.6]$ filter
combination.

\section*{Color-Magnitude Relation}	

As an illustration of the overall depth of the FLS, and the expected colors of early-type galaxies,
Figure~\ref{fig:CMR} shows the $R-[3.6]$ color-magnitude diagram for all galaxies 
in the 4 degree$^{2}$ Main Field.  The IRAC component
of the FLS totals 60s of coverage, reaching a survey depth 
of about $[3.6]=18.5$ on the Vega magnitude scale. The R-band component was obtained at the Kitt Peak 4m Mayall
Telescope (Fadda et al. 2004).

The solid lines show the
expected colors of early-types from models generated using the code of Bruzual and Charlot (2003).
The slight slope is a reflection of the well-known color-mass-metallicity effect; 
fainter, less massive galaxies appear bluer in color. 

%$M_{\star}$, the characteristic magnitude at each redshift, is denoted by an asterisk.

%\begin{figure}
%     \centering \leavevmode
%         \epsfxsize=.8\textwidth \epsfbox{wilsong_fig1.ps}
%     \caption{Blah, blah}
%     \end{figure}

\section*{Photometric Redshift Accuracy}

Figure~\ref{fig:photoz} shows the photometric redshift (inferred from the
$R-[3.6]$ color) versus the spectroscopic redshift, for each of 26 clusters for which
a spectroscopic redshift of a red-sequence galaxy was available. The key shows the source
of each of the spectra. The large symbols indicate
clusters for which more than one spectroscopic redshift was measured.

The photometeric and spectroscopic redshifts are in excellent agreement,
with an r.m.s. dispersion of $\delta z = 0.07$. 
This is an accuracy comparable to the best four (or more) passband photometric studies. 
The advantage here is, of course,
that one is determining a photometric redshift for each cluster by calculating a mean color, 
averaging over many early-type galaxies with intrinsically similar color at the same redshift. 

%\begin{figure}
%     \centering \leavevmode
%         \epsfxsize=.8\textwidth \epsfbox{wilsong_fig1.ps}
%     \caption{Blah, blah}
%     \end{figure}

%begin{figure}
%     \plotone{wilsong_fig1.ps}
%     \caption{Blah, blah}
%     \end{figure}

\section*{Cluster Examples}
	
\begin{figure}[!ht]
\begin{center}
\scalebox{0.9}{
  \includegraphics{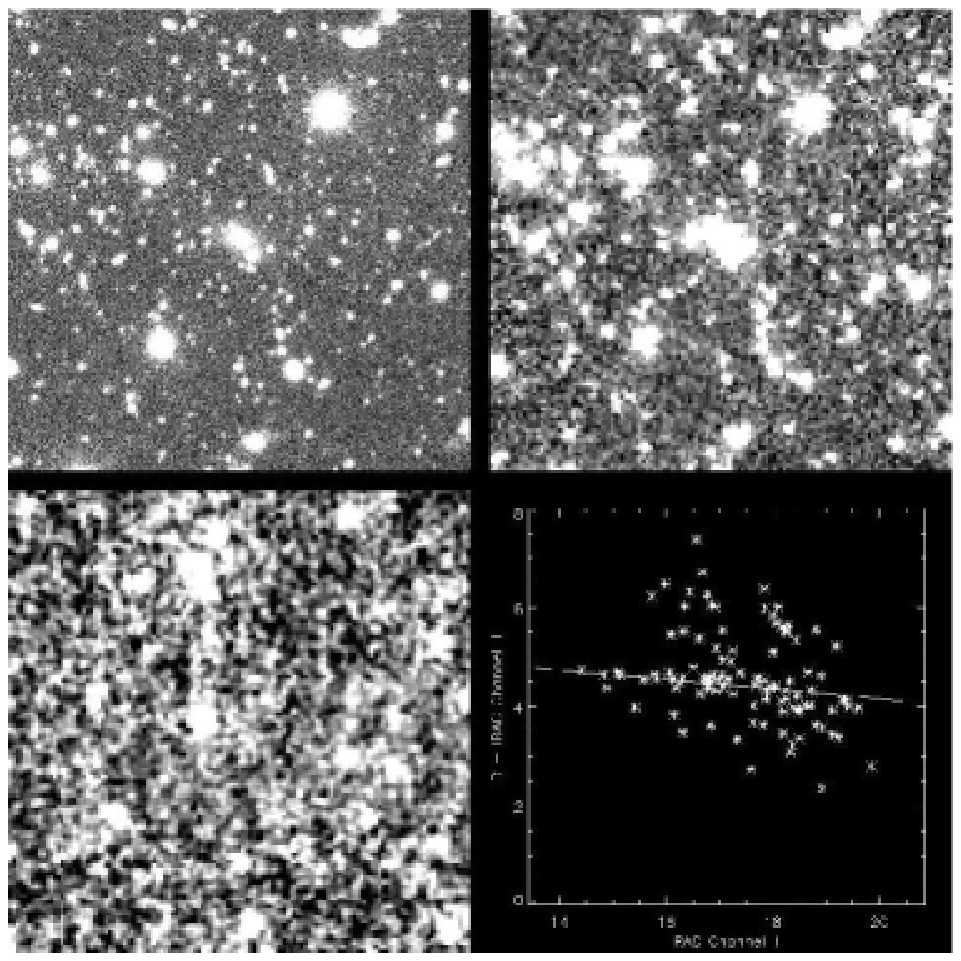}	
}
\caption{
A cluster at $z=0.55$ detected in the First Look Survey using the CRS technique.
Clockwise from upper left the four panels show the $R$ image, the $[3.6]$ image, the $R-[3.6]$ color-magnitude diagram, and the MIPS $[24]$ image. The f.o.v. is 500 kpc at the cluster redshift.}
 \label{fig:FLScluster55}
\end{center}

\begin{center}
\scalebox{0.9}{
 \includegraphics{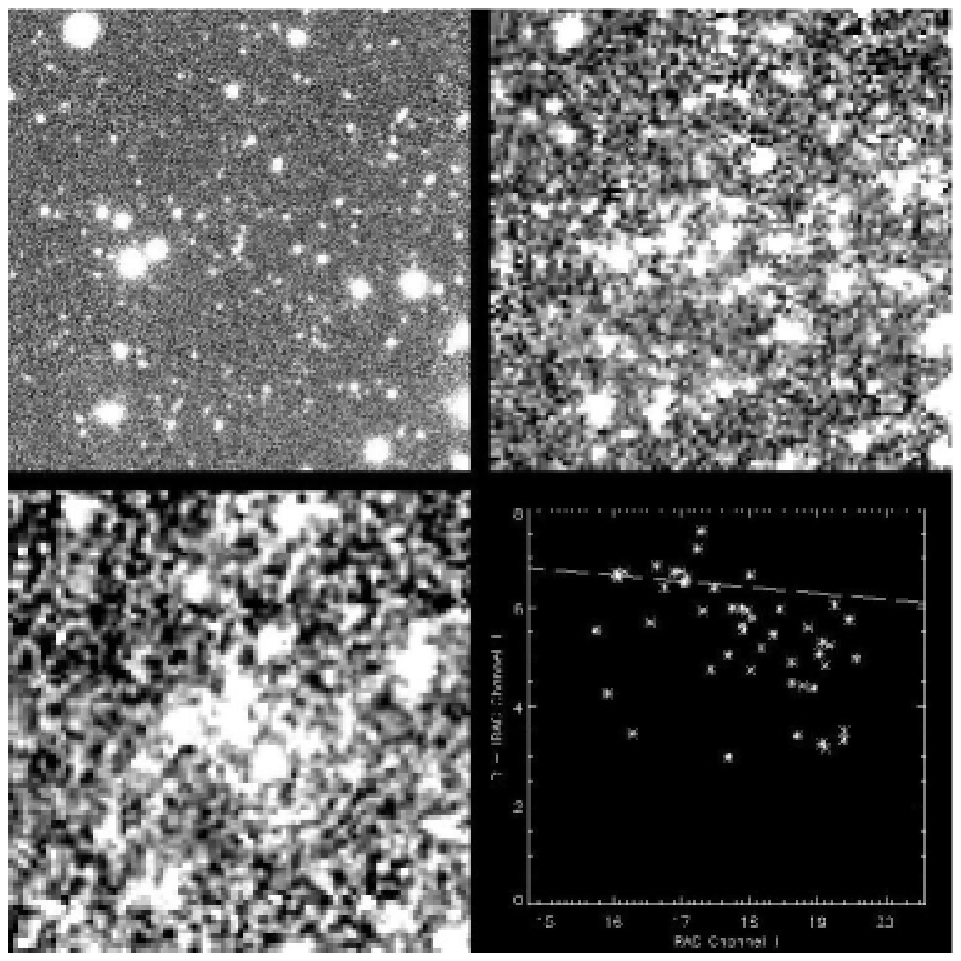}	
}
\caption{
As for Figure 3, but for  a cluster at $z=1.12$. 
}
 \label{fig:FLScluster}
\end{center}
\end{figure}

Figure~\ref{fig:FLScluster55} shows an example of a cluster detected at $z=0.55$.
Clockwise from upper left the four panels show the $R$ image, the $[3.6]$ image, the $R-[3.6]$ color-magnitude diagram,
and the MIPS $[24]$ image. 

For comparision, Figure~\ref{fig:FLScluster} shows an example of a cluster at $z=1.12$.
{\em Spitzer} will be able to detect clusters to even higher redshift using deeper datasets and
alternate filter combinations.
In collaboration with the SWIRE Legacy Team we are now applying
the CRS technique to the $50$ degree$^2$ SWIRE Legacy Fields.

\section*{Further Reading}

This paper presented an IR adaptation of the CRS technique for detecting clusters
at high redshift. More details about the IRAC component of the FLS
may be found in Lacy et al. (2005). A study of the evolution of the $3.6 \micron$ cluster luminosity function may be found in
Muzzin et al. (this volume) and Muzzin et al. (2005a), and a study of cluster $24 \micron$ sources
in Muzzin et al. (2005b).

%Muzzin et al. (2005b) (a study of the $24 \micron$ sources).
%Muzzin et al. (this volume), Muzzin et al. (2005a) and
%Muzzin et al. (2005b).

%\subsection{}   %%% Second level section head (remove "%" symbol)
%\subsubsection{}   %%% Lowest level section head (remove "%" symbol)
%\section*{}	%%% Unnumbered top level section head (remove "%" symbol)
%\subsection*{}   %%% Unnumbered second level section head (remove "%" symbol)

%\acknowledgements %%% Text of acknowledgements runs on after this command.

%%% THE BIBLIOGRAPHY
%%%
%%% CONSULT SECTION 3 OF "INSTRUCTIONS FOR AUTHORS" FOR HOW TO USE NATBIB.
%%% AUTHORS ARE ENCOURAGED TO USE EITHER THE "THEBIBLIOGRAPY" ENVIRONMENT
%%% BY UNCOMMENTING (DELETING THE "%" SYMBOL) THE COMMANDS BELOW, OR BY
%%% USING THE BIBTEX ENVIRONMENT. TO FIND OUT WHICH IS APPLICABLE TO YOUR
%%% CONTRIBUTION, CONSULT THE VOLUME EDITORS FOR YOUR PROCEEDINGS.
%%%

\end{document}